\documentclass[aps,tightenlines,preprint]{revtex4}
\usepackage{epsfig}
\newcommand{\n}{\noindent}
\begin{document}

\title{Level velocity statistics of hyperbolic chaos}

\author{R. Sankaranarayanan \footnote{Present Address: Center for Nonlinear
Dynamics, Department of Physics, Bharathidasan University, Trichy 620024,
India.}} 
\affiliation{Physical Research Laboratory,\\ Navrangpura, Ahmedabad 380009,
India.} 
%\date{\today}

\begin{abstract}
A generalized version of standard map is quantized as a model of quantum
chaos. It is shown that, in hyperbolic chaotic regime, second moment of
quantum level velocity is $\sim 1/\hbar$ as predicted by the random matrix
theory. 
\end{abstract}
\maketitle

\section{Introduction}

Chaotic systems are characterized by positive lyapunov exponents such that
in phase space neighbourhood trajectories diverge exponentially. Among the
chaotic systems there are special class of systems which are {\it completely}
chaotic or hyperbolic \cite{ott}. Phase space of hyperbolic systems has only
unstable orbits as there are expanding and contracting real directions with
positive and negative lyapunov exponents respectively. Area preserving maps
like cat map and baker map are examples of hyperbolic systems. Even the well
known standard map of kicked rotor, a text book paradigm of Hamiltonian chaos
\cite{lich}, is not proven to be hyperbolic even for strong external kick
strength. 

In the study of chaotic quantum systems, it is of fundamental interest to
characterize highly chaotic (but not known to be hyperbolic) and hyperbolic
chaotic regimes in quantum domain. This forms our motivation here, and to
pursue further it would be more appropriate to quantize a {\it single}
dynamical system which has parameters for highly chaotic {\it and}
hyperbolic chaotic regimes. In one of our earlier works, a generalized
version of standard map was introduced to study the dynamics of a
kicked particle trapped inside an one dimensional infinite square well
potential \cite{sankar01}. The generalization, arising from length scales
namely well width and field wave length, has parameters to fulfill the
present requirement. With our knowledge, there is no other single system
possessing parameters for the above mentioned classical regimes. 

In quantum domain, dynamics of levels in parameter space is known to have
manifestations of classical complexity. While quantum levels cross each other
for regular case, they exhibit avoided crossings when underlying classical
dynamics is chaotic. Level dynamics can be described by level velocity wherein
system parameter plays the role of pseudo-time. In Ref. \cite{gaspard} the
notion of curvature i.e., second derivative of levels with parameter, is
introduced to quantify the avoided crossings. It is known that the curvature
distribution of chaotic system follows an universal behaviour \cite{models}.

In this connection, one another quantity of importance is the second moment
of level velocity. A semiclassical analysis of kicked system shows that
second moment is the sum of all classical time correlations of the kicking
potential, such that lowest (zeroth) order correlation being the Random Matrix
Theory (RMT) predicted second moment \cite{arul99}. It is also known for
quantized standard map that, in highly chaotic regime there are systematic
deviations between second moment and the corresponding RMT prediction. The
deviations are thus related to the non-vanishing higher order classical time
correlations \cite{arul99}. 

In the present work, we introduce quantum version of a generalized standard
map as a model of quantum chaos to study the level velocity statistics. In
particular, we compute the second moment and compare with the RMT prediction
in order to characterize different chaotic regimes.

\section{The Model}
\subsection{Classical system}

Considering a particle trapped in a one dimensional infinite square well 
potential $V_0(q)$ of unit width (hard walls at $q=\pm 1/2$), which 
experiences a periodic kick from an external pulsed field. The Hamiltonian is 

\begin{equation}
\tilde {H} = {p^2\over 2} + V_0(q) + {k\lambda\over 4{\pi}^2} 
\cos\left({2\pi q\over\lambda}\right) \sum_j \delta (j-t)
\end{equation}

\n and underlying kick to kick dynamics of the particle is {\it equivalent}
to discrete dynamics described by a dimensionless area-preserving mapping:

\begin{eqnarray}
p_{j+1}&=&p_j+{k\over 2\pi}\sin(2\pi rq_j ) \nonumber \\  
q_{j+1}&=&q_{j}+p_{j+1}
\label{gsm}
\end{eqnarray}

\n which is defined on 2-torus i.e., a unit square $[-1/2,1/2)\times 
[-1/2,1/2)$ with periodic boundaries. Here $r=1/\lambda$ is ratio of two
length scales of the system namely, well width and field wavelength;
$k$ is effective strength of the kick. This is the {\it Generalized Standard
Map} (GSM) which was introduced in our earlier studies on the above
Hamiltonian \cite{sankar01}.

GSM is continuous when $r$ is integer and discontinuous otherwise. One can
immediately recognize that widely studied standard map of kicked rotor is a
special case ($r=1$) of GSM. Since the standard map is a continuous map,
for small $(k<1)$ dynamics is predominantly regular wherein many rotational
invariant circles (also called as KAM tori) are interspersed in the phase
space. They act as forbidden barriers for chaotic orbits to diffuse. Gradual
destruction of these invariant structures with the increase of $k$, leads to
onset of chaos; for $k\gg 1$, dynamics is highly chaotic. On the other hand,
when $r$ is non-integer no KAM tori exist in the phase space. In this case,
depends of the parameter $r$ the phase space is either mixed or fully chaotic
even for small $k$ values.

The Jacobian ${\bf J}$ of GSM is such that

\begin{equation}
|\hbox{Trace}\;{\bf J}| = |2+kr\cos(2\pi rq_j)| \, .
\end{equation}

\n Since $|q_j|\le 1/2$, for $r\le 1/2 \;\; |\hbox{Trace}\;{\bf J}| > 2$.
That is to say Jacobian has real eigenvalues. In other words, the system is
{\it completely} chaotic or hyperbolic for $r\le 1/2$. In this regime there
are contracting and expanding real directions or alternatively stable and
unstable manifolds throughout the phase space. Thus GSM is realized as a
rare class of dynamical system as it has parameters for both highly chaotic
and hyperbolic chaotic regimes. 

\subsection{Quantum system}

GSM arises from the equation of motion of free particle in presence of a field
$V(q)$ which is applied as time periodic impulse. The field is defined as:
$V(q) = k\cos(2\pi rq)/(4{\pi}^2r); V(q) = V(q+1)$ and the Hamiltonian is 

\begin{equation}
H = {p^2\over 2} + V(q)\sum_j \delta (j-t) \, .
\label{ham}
\end{equation}

\n By integrating Shr\"{o}dinger equation over unit time we obtain
corresponding quantum propagator as

\begin{equation} 
\hat{U} = e^{-i{\hat{p}}^2/2\hbar} \,\,\, e^{-iV(\hat{q})/\hbar} \, .
\end{equation}

\n Then the quantum dynamics can be described as $|\psi (t+1)\rangle =
\hat{U}| \psi (t)\rangle$, which is a quantum analogue of the classical map.

On quantizing 2-torus phase space by introducing periodic boundary conditions
both in $q$ and $p$ \cite{ford91} we have: $\hat{q}|n\rangle = (n/N)|n\rangle
\; ; \; \hat{p}|m\rangle = (m/N)|m\rangle$ where $n,m = -N/2, -N/2+1\ldots
N/2-1$. Here $N={(2\pi\hbar)}^{-1}$ is the dimensionality of the Hilbert space
and the semiclassical limit is $N\rightarrow \infty$. The position and
momentum eigenstates obey the periodicity $|n+N\rangle = |n\rangle\; ;\;
|m+N\rangle = |m\rangle$ and the transformation function is 

\begin{equation}
\langle n|m\rangle = {1\over \sqrt{N}} 
\exp\left[{i2\pi mn\over N}\right] \, .
\end{equation}

\n Being a homogeneous linear system, Shr\"{o}dinger equation in finite $N$
dimensional space has solutions $|\phi_j\rangle ,\; j=1,2,\ldots N$, which
are linearly independent. Since the Hamiltonian is time periodic (unit period),
according to Floquet theory \cite{jor} the solutions satisfy eigenvalue
equation $\hat{U}|\phi_j\rangle = e^{-i\phi_j}|\phi_j\rangle$. Eigenstates
$|\phi_j\rangle$ are quasienergy states and eigenangles $\phi_j$ are 
quasienergies. Then general solution at a given time is $|\psi\rangle =
\sum_j c_j|\phi_j\rangle$ where $c_j$ are constants. As a consequence of
hermiticity of Hamiltonian, quasienergy states are orthogonal and they form
a complete set in finite $N$ dimensional space.

Matrix form of the propagator in discrete position representation 
is \cite{arul97} 

\begin{equation}
U_{nn^\prime} \equiv \langle n|\hat{U}|n'\rangle = {1\over \sqrt{N}}
\exp\left[ -i\pi \left\{ {1\over 4} - {{(n-n')}^2\over N} + 
2NV\left({n'\over N}\right) \right\} \right] \, .
\end{equation}

\n The Hamiltonian (\ref{ham}) has reflection symmetry about the origin
i.e., $H(q,p) = H(-q,-p)$. This symmetry is reflected in the quantum
propagator matrix 

\begin{equation}
U_{nn^\prime} = {e^{-i\pi/4}\over\sqrt{N}}\; e^{i\pi{(n-n')}^2/N} 
\exp\left\{{-ikN\over 2\pi r} \cos \left[{2\pi r\over N}(n'+\alpha)
\right]\right\}
\end{equation}

\n through the relation $[\hat{U},\hat{R}]=0$ where the hermitian operator
$\hat{R}$ is defined as 

\begin{equation}
\begin{array}{llll}
\hat{R}|n\rangle &=& |-n\rangle &\hbox{for} \;\;\alpha = 0 \\
&=& |-n-1\rangle &\hbox{for} \;\;\alpha = 0.5 \, .
\end{array}
\end{equation}
 
\n Note that a phase factor $\alpha$ is introduced in the matrix element to
avoid exact quantum symmetry. Since $\hat{R}^2 = 1$ we may label the
eigenstates of $\hat{U}$ with eigenvalues $\pm 1$ of $\hat{R}$ i.e., the
states are $|\phi_\pm\rangle$. For $\alpha = 0.5$, symmetry matrix of order
$N$ is 

\begin{equation}
R_N = \langle n|\hat{R}|n^\prime \rangle = 
\delta (n+n^\prime +1) \;\;\;\;\; (\hbox{mod}\; N)
\end{equation}

\n which has ones along secondary diagonal and zeros elsewhere. Then the state 
components have a relation $\langle -n-1|\phi\rangle = \pm\langle n
|\phi\rangle$ i.e., 

\begin{equation}
|\phi_\pm\rangle = \left(\begin{array}{r} |v\rangle \\ 
\pm R_{N/2}|v\rangle \end{array}\right) \, . 
\end{equation}

Eigenstates can be numerically obtained by diagonalizing the matrix
$U_{nn^\prime}$ of order $N$. If $N$ is even integer, there are $N/2$ even
parity states \{$|\phi_+\rangle$\} and $N/2$ odd parity states
\{$|\phi_-\rangle$\}. On exploiting $R$-symmetry, the diagonalization can be
reduced to the matrix of order $N/2$ by standard procedure \cite{sar89}.
The reduced matrix is 

\begin{eqnarray}
{\cal U}_{nn^\prime} &=& {e^{-i\pi/4}\over\sqrt{N}} 
\exp\left\{{-ikN\over 2\pi r} \cos \left[ {2\pi r\over N} \left(n+{1\over 2}-
{N\over 2}\right) \right]\right\} \nonumber \\[8pt] 
&& \;\;\;\;\;\;\;\;\;\;\;\;\;\;\; \times \left\{e^{i\pi{(n-n')}^2/N} 
\pm e^{i\pi{(n+n'+1)}^2/N} \right\}
\end{eqnarray}

\n where $n,n^\prime = 0,1,\; .\; .\; .\;N/2-1$. Now the separation of parity
states is obvious. 

\section{Spectral statistics}

\begin{figure}[h]
\centerline{\psfig{figure=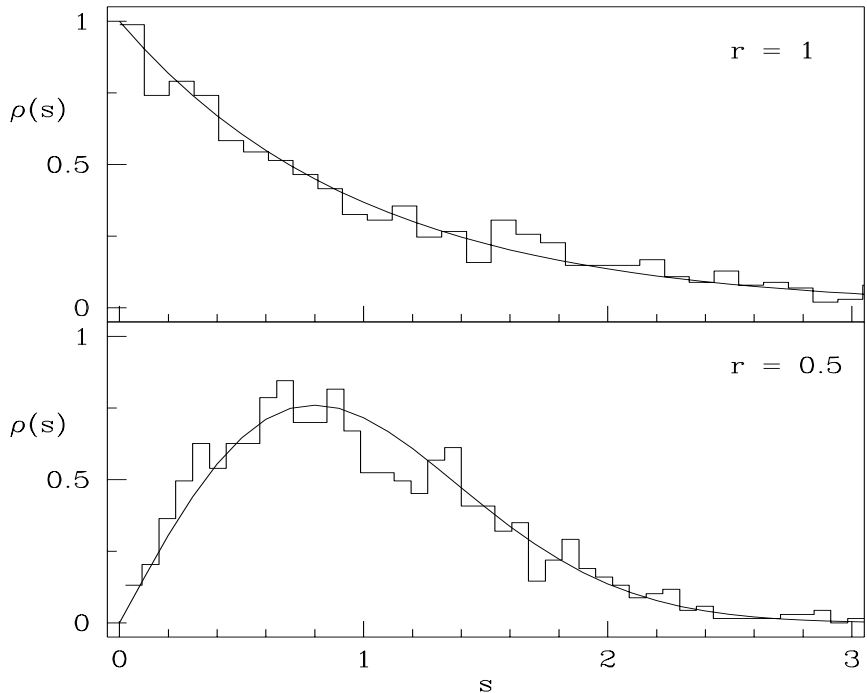,height=10cm,width=12cm}}
\caption{Nearest neighbour spacing distributions of 1000 quasienergies which
correspond to even parity states for $k=0.3$ with $\alpha = 0.5$. Smooth
curves are the Poisson and Wigner distribution.}
\label{nns}
\end{figure}

One of the standard statistical measures for a chaotic quantum system is the 
nearest neighbour spacing distribution of quantum levels. For a regular system,
levels are clustered such that the spacings follow Poisson distribution. On
the other hand, in chaotic case the levels exhibit repulsion such that spacings
exhibit RMT predicted Wigner distribution. We expect the classical complexity
of GSM, arises due to the parameter $r$, will also have manifestation in
the spectral spacings. It is evident from Fig. \ref{nns} that, in contrast 
to the predominantly regular case, for $r=0.5$ (hyperbolic) the spacings
follow Wigner distribution. 

\subsection{Level velocities}

Having seen the effect of $r$ in the spectral spacing, we further investigate
on dynamics of the quasienergies in parameter space i.e., level velocities.
To be specific, we study second moments of level velocities in different
classical regimes. We take $\alpha = 0.35$ so that $R$-symmetry is broken in
the quantum system, and the factor $\alpha$ is dropped out in following
expressions for the sake of convenience. Quasienergies $\phi_j \equiv
\phi_j(k,r)$ have scaled velocities:

\begin{eqnarray} 
x_j &=& \left({2\pi^2 r^2\over N}\right) {\partial\phi_j\over\partial k} 
\nonumber \\[8pt]
&=& \left({2\pi^2 r^2\over N}\right) \left[ {1\over\hbar } 
\langle\phi_j|\partial V/\partial k|\phi_j\rangle \right] \nonumber \\[8pt]
&=& \pi r\sum_n \cos (2\pi rn/N) {|\langle n|\phi_j\rangle |}^2
\end{eqnarray}

\n and 

\begin{eqnarray} 
y_j &=& \left({2\pi r^2\over Nk}\right) {\partial\phi_j\over \partial r} 
\nonumber \\[8pt]
&=& \left({2\pi r^2\over Nk}\right) \left[ {1\over\hbar} 
\langle\phi_j|\partial V/\partial r|\phi_j\rangle \right] \nonumber \\[8pt]
&=& -2\pi r\sum_n (n/N)\sin (2\pi rn/N) {|\langle n|\phi_j\rangle |}^2 - 
\sum_n \cos (2\pi rn/N) {|\langle n|\phi_j\rangle |}^2 .
\end{eqnarray}

\n Average velocities in semiclassical limit are 

\begin{eqnarray}
\langle x\rangle &=& {1\over N}\sum_j x_j = \sin(\pi r) \nonumber \\
\langle y\rangle &=& {1\over N}\sum_j y_j = \cos(\pi r) - 
{2\sin(\pi r)\over\pi r} \, .
\end{eqnarray}

Then the second moment of $x$ is given by

\begin{eqnarray}
\langle x^2 \rangle &=& {1\over N} \sum_j x_j^2 \nonumber \\[8pt] 
&=& {(\pi r)}^2 \left\{\sum_n \cos^2 (2\pi rn/N)
\left\langle {|\langle n|\phi_j\rangle|}^4\right\rangle \right .
\nonumber \\[8pt]
&& ~~~~~~~~~~~ + \left . \sum_{n\neq n'} \cos (2\pi rn/N)
\cos (2\pi rn^\prime/N) \left\langle {|\langle n|\phi_j\rangle|}^2 
{|\langle n'|\phi_j\rangle|}^2 \right\rangle \right\}
\label{x2}
\end{eqnarray}

\n Assuming that spectral averaged eigenfunction components are independent
of specific position eigenvalues $n$, terms within the angle brackets can be
taken out of the sum. Then

\begin{eqnarray}
\langle x^2\rangle &\sim& {(\pi r)}^2 \left\{\left[\left\langle 
{|\langle n|\phi_j\rangle|}^4 \right\rangle - \left\langle 
{|\langle n|\phi_j\rangle|}^2 {|\langle n'|\phi_j\rangle|}^2 
\right\rangle \right] \sum_n \cos^2(2\pi rn/N) \right . \nonumber \\[8pt]
&& ~~~~~~~~~~~+ \left . \left\langle {|\langle n|\phi_j\rangle|}^2 
{|\langle n'|\phi_j\rangle|}^2 \right\rangle {\left[\sum_n 
\cos (2\pi rn/N)\right]}^2 \right\} \, .
\end{eqnarray}

\n In chaotic regimes, standard RMT results \cite{brody81}

\begin{equation}
\begin{array}{rll}
\left\langle{|\langle n|\phi_j\rangle|}^4 \right\rangle &=&
3{[N(N+2)]}^{-1} \simeq 3N^{-2} \nonumber \\[8pt]
\left\langle{|\langle n|\phi_j\rangle|}^2{|\langle n'|\phi_j\rangle|}^2 
\right\rangle &=& {[N(N+2)]}^{-1} \simeq N^{-2} 
\end{array}
\label{rmt}
\end{equation}

\n which correspond to Gaussian orthogonal ensemble are applicable here as 
well. It should be noted that application of RMT results essentially adopt
the assumption made above. Replacing the sum by integration in semiclassical
limit we arrive to

\begin{equation}
{\langle x^2\rangle }_{\hbox{\tiny RMT}} = {{(\pi r)}^2\over N} 
\left[1 + {\sin (2\pi r)\over 2\pi r}\right] + {\langle x\rangle}^2 \, .
\label{x2_rmt}
\end{equation}

\n Similarly the second moment of $y$ is 

\begin{eqnarray}
\langle y^2 \rangle &=& {1\over N} \sum_j y_j^2 \\[8pt]
&=& \sum_n \{{\left[2\pi r(n/N) 
\sin (2\pi rn/N)\right]}^2 + \cos^2 (2\pi rn/N) \nonumber \\[8pt] 
&& + \; 4\pi r(n/N)\sin(2\pi r n/N)\cos (2\pi rn/N)\}
\left\langle {|\langle n|\phi_j\rangle|}^4\right\rangle \nonumber \\[8pt]
&& + \sum_{n\neq n^\prime} \{{(2\pi r/N)}^2nn^\prime\sin (2\pi rn/N)
\sin (2\pi rn^\prime /N) + \cos (2\pi rn/N) \cos (2\pi rn^\prime /N)
\nonumber \\[8pt] 
&& + \; 4\pi r(n/N)\sin (2\pi rn/N)\cos(2\pi rn^\prime /N)\} \left\langle
{|\langle n|\phi_j\rangle|}^2 {|\langle n'|\phi_j\rangle|}^2 \right\rangle
\label{y2}
\end{eqnarray}

\n and the RMT prediction is 

\begin{equation}
{\langle y^2\rangle }_{\hbox{\tiny RMT}} = {1\over N} \left\{ 1 + 
{{(\pi r)}^2\over 3} + {\sin (2\pi r)\over 2\pi r} \left[{5\over 2}-
{(\pi r)}^2\right] -{3\over 2}\cos (2\pi r)\right\} + {\langle y\rangle}^2\, .
\label{y2_rmt} 
\end{equation}

In chaotic regime, quantum states are such that the quantities in left hand
side of Eqn. (\ref{rmt}) fluctuate about the respective RMT values. These
fluctuations could lead to the failure of RMT in predicting the second moment.
In Ref. \cite{arul99} a semiclassical analysis on the systematic deviation
between the second moment and its RMT prediction has been made. Here we are
interested to see the validity of the RMT prediction in different classical
regimes. For that we define normalized deviations:

\begin{equation}
\Delta_x = \left|{\langle x^2 \rangle - {\langle x^2\rangle}_{\hbox{\tiny RMT}}
\over\langle x^2\rangle}\right| \;\; ; \;\;
\Delta_y = \left|{\langle y^2 \rangle - {\langle y^2\rangle}_{\hbox{\tiny RMT}}
\over\langle y^2\rangle}\right| 
\end{equation}

\n and taking average of the two positive quantities as 

\begin{equation} 
\Delta = {\Delta_x+\Delta_y\over 2} \, .
\end{equation}

\begin{figure}[h]
\centerline{\psfig{figure=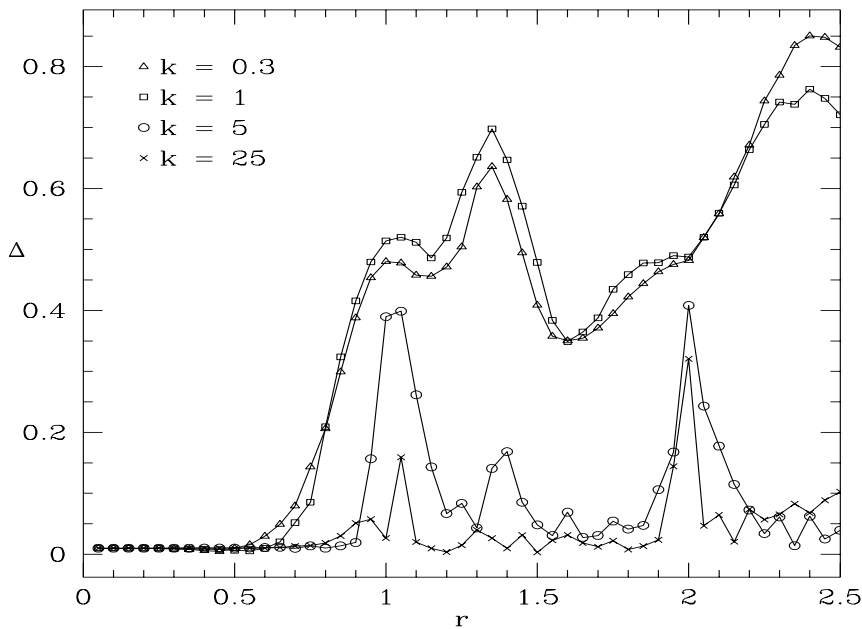,height=9cm,width=12cm}}
\caption{Deviation between level velocity second moments and RMT predictions.
In all the cases we have taken $\alpha = 0.35$ and $N=200$.}
\label{dev1}
\end{figure}

The deviation is calculated for various parameters and plotted in
Fig. \ref{dev1}. Let us first consider the data obtained for small values of
$k$ (0.3 and 0.1). As we see the deviation is nearly zero for $r\le 1/2$.
For further values of $r$, the deviation is found to be significantly large.
This is expected as the underlying classical dynamics is predominantly regular
or mixed and the RMT is not applicable. On the other hand, the data for large
$k$ (5 and 25) exhibit different behaviour. Though the classical phase space
is highly chaotic for these parameters, significant deviation observed for
various values of $r$ shows the failure of RMT prediction. However, it is
noticable from the four sets of observation that for $r \le 1/2$ the deviation
is nearly zero.  

\begin{figure}
\centerline{\psfig{figure=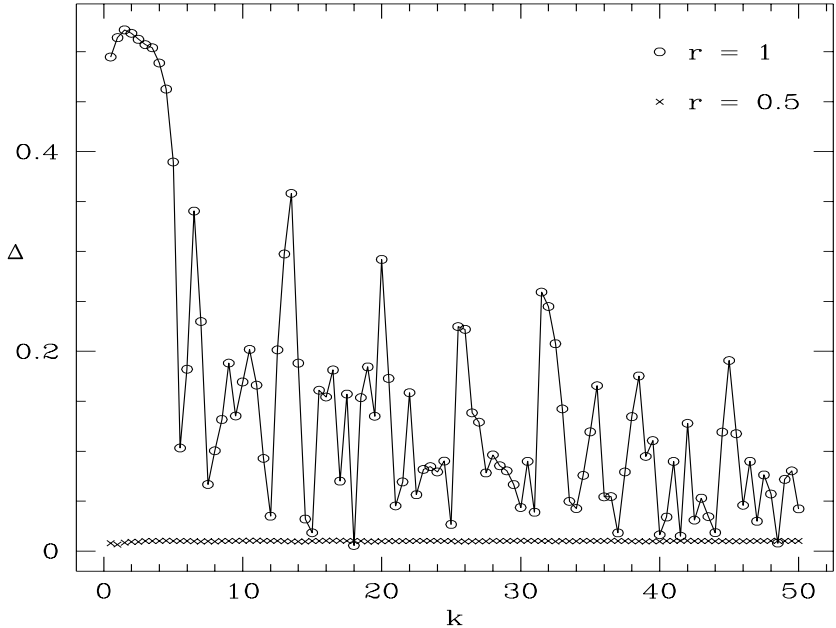,height=9cm,width=12cm}}
\caption{Deviation between level velocity second moments and RMT predictions.
We have taken $\alpha = 0.35$ and $N=200$.}
\label{dev2}
\end{figure}

In Fig. \ref{dev2}, the deviation is plotted for two cases by varying $k$.
For $r=1$ the deviation has strong fluctuations with $k$. The large 
deviation for $k<5$ is due to the predominantly regular/mixed behaviour of
the underlying classical system. We also observe significant deviation when
$k$ is close to integer multiples of $2\pi$. This may be attributed to the
presence of ``accelerator modes'' \cite{zas97}, which are small regular regions
embedded in the sea of chaotic phase space. On the other hand, for $r=0.5$
the deviation remains negligible irrespective of $k$ values. This shows that
RMT predicts second moment of level velocity in hyperbolic chaotic regime. From
the Eqn. (\ref{x2_rmt}), we see that $\langle x^2 \rangle_{\hbox{\tiny RMT}}
\sim 1/N$ or $\langle {(\partial\phi/\partial k)}^2 \rangle_{\hbox{\tiny RMT}}
\sim N \sim 1/\hbar$. We obtain the same result for the velocity with respect
to $r$ as well. 

\subsection{Sawtooth map}

In this section we study another quantum model whose classical counterpart
has hyperbolic regime. Consider a free particle that is subjected to a time
periodic impulsive potential $V(q)=-\lambda q^2/2;V(q+1)=V(q)$. Kick to kick
dynamics of such a particle is described by the sawtooth map \cite{cary81}:

\begin{equation}
\begin{array}{lll}
p_{j+1} &=& p_j + \lambda q_j \\ q_{j+1} &=& q_j + p_{j+1} 
\end{array}
\end{equation}

\n defined on a unit torus. This map is stable for $-4 < \lambda < 0$ and
unstable (or hyperbolic) otherwise. Quantized version of this map can be
obtained, as usual, upon introducing periodic boundaries in $q$ and $p$.
Details of quantized sawtooth map are given elsewhere \cite{arul95}. Note
that here also the earlier symmetry arguments hold. From the corresponding
quasienergies and quasienergy states, we define the scaled level velocity as 

\begin{equation}
z_j = \left({-1\over N\pi}\right){d\phi_j\over d\lambda}
= -2\left\langle\phi_j\left|{dV\over d\lambda}\right|\phi_j\right\rangle
= \sum_n {(n/N)}^2 {|\langle n|\phi_j\rangle|}^2 
\end{equation} 

\n with average $\langle z\rangle = 1/12$. Then the second moment is given by

\begin{equation}
\langle z^2\rangle = {1\over N} \sum_j z_j^2 = \sum_n {(n/N)}^4
\left\langle{|\langle n|\phi_j\rangle|}^4\right\rangle + \sum_{n\neq n^\prime} 
{(n/N)}^2 {(n^\prime/N)}^2 \left\langle {|\langle n|\phi_j\rangle|}^2
{|\langle n^\prime|\phi_j\rangle|}^2 \right\rangle \, .
\label{z2}
\end{equation}

\n Repeating the earlier procedures, we find the RMT prediction as 

\begin{equation}
{\langle z^2\rangle}_{\hbox{\tiny RMT}} = {1\over 40N} + 
{\langle z\rangle}^2 \, .
\label{z2_rmt}
\end{equation}

\n Defining normalized deviation as 

\begin{equation}
\Delta '= \left|{\langle z^2 \rangle - {\langle z^2\rangle}_{\hbox{\tiny RMT}}
\over\langle z^2\rangle}\right|
\end{equation}

\begin{figure}
\centerline{\psfig{figure=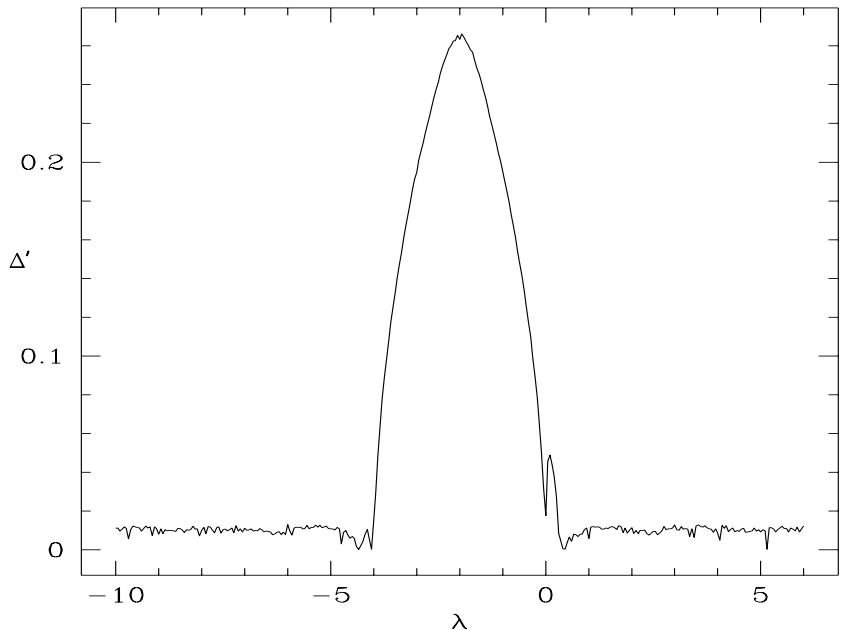,height=9cm,width=12cm}}
\caption{Deviation $\Delta '$ for the quantum sawtooth map with $\alpha =
0.35$ and $N=200$.}
\label{saw}
\end{figure}

\n in Fig. \ref{saw} we have plotted the deviation for different $\lambda$
values. In the stable region $-4<\lambda<0$, where the RMT result is not
applicable, the deviation is large. Evidently, RMT predicts level velocity
second moment in hyperbolic regime. Eqn. (\ref{z2_rmt}) shows that $\langle
z^2 \rangle_{\hbox{\tiny RMT}} \sim 1/N$ or $\langle {(d\phi/d\lambda)}^2
\rangle_{\hbox{\tiny RMT}} \sim N \sim 1/\hbar$. Here also we see that second
moment is inversely proportional to the Planck constant for hyperbolic chaos.

\section {conclusion}

In this paper we have introduced quantum version of a generalized standard
map. The classical system corresponds to the dynamics of a particle, trapped
in an 1D infinite square well potential, in presence of time-periodic kicks.
For different classical regimes, the second moment of quantum level velocity
is computed and compared with the RMT prediction. We have shown that while
the prediction fails in highly chaotic regime, the second moment is well
predicted by the RMT as $\sim 1/\hbar$ in hyperbolic chaotic regime. By
considering another example viz., quantum sawtooth map, the RMT prediction
of second moment in hyperbolic chaotic regime has been reinforced. We hope
the presented result would shed new lights to explore more on quantum
hyperbolic chaos. \\

\n {\bf Acknowledgement} \\

The author is greatful to Dr. A. Lakshminarayan for useful discussions.

\end{document}